\documentclass{aa}
\usepackage{txfonts}
\usepackage{graphicx}
\usepackage{natbib}
\usepackage{amssymb}
\bibpunct{(}{)}{;}{a}{}{,}
\begin{document}
   \title{SiO in C-rich circumstellar envelopes of AGB stars: effects of non-LTE chemistry and grain adsorption}
 
   \titlerunning{SiO in C-rich circumstellar envelopes of AGB stars}

   \author{F.~L. Sch\"oier\inst{1} \and H. Olofsson\inst{1,2} \and A.~A. Lundgren\inst{3}}

   \offprints{F.~L. Sch\"oier \\ \email{fredrik@astro.su.se}}

   \institute{Stockholm Observatory, AlbaNova University Center, SE-106 91 Stockholm, Sweden
   \and Onsala Space Observatory, SE-439 92 Onsala, Sweden          
   \and European Southern Observatory, Casilla 19001, Santiago 19, Chile}
   \date{Received; accepted}

   \abstract{New SiO multi-transition millimetre line observations of a sample of carbon stars, including $J$$\,=\,$8$\rightarrow$7 observations with the APEX telescope, are used to probe the role of non-equilibrium chemistry and the influence of grains in circumstellar envelopes of carbon stars. A detailed radiative transfer modelling, including the effect of dust emission in the excitation analysis, of the observed SiO line emission is performed. A combination of low- and high-energy lines are important in constraining the abundance distribution. It is found that the fractional abundance of SiO in these C-rich environments can be several orders of magnitude higher than predicted by equilibrium stellar atmosphere chemistry. 
In fact, the SiO abundance distribution of carbon stars closely mimic that of M-type (O-rich) AGB stars.     
A possible explanation for this behaviour is a shock-induced chemistry, but also the influence of dust grains, both as a source for depletion as well as production of SiO, needs to be further investigated.  As observed for M-type AGB stars, a clear trend that the SiO fractional abundance decreases as the mass-loss rate of the star increases is found for the carbon stars. This indicates that SiO is accreted onto dust grains in the circumstellar envelopes.

   \keywords{Stars: AGB and post-AGB -- Stars: carbon -- Stars: circumstellar matter -- Stars: late-type -- Stars: mass-loss}
   }
   \maketitle
%

\section{Introduction}
In addition to dictating the evolutionary time scale, stellar winds from late-type stars are thought to be important for the enrichment of heavy elements in the ISM \citep{Gustafsson04}. The stellar wind carries the results of internal nuclear processes, activated during evolution along the asymptotic giant branch (AGB), and thus contributes to the chemical 
evolution of galaxies.

Red giant stars located on the AGB divide into two chemically distinct groups; carbon stars (C/O$>$1), which show a more rich chemistry than the M-type (O-rich) stars (C/O$<$1).  
The molecular- and grain-type setups in the circumstellar envelopes (CSEs) around these mass losing stars are to a large extent determined by the 
C/O-ratio of the central star.
Of paramount importance 
is to know how much of the elements that are eventually locked in dust particles and how much remains in the gas phase.

Observations of SiO line emission is particularly interesting because it is potentially a
useful probe of the formation and evolution of dust grains in CSEs, as
well as CSE dynamics: hinted at from early interferometric observations \citep{Lucas92,Sahai93}. A major survey
of SiO emission from  M-type AGB stars
was performed by \citet{Delgado03b}. Supplemented by a detailed numerical
modelling of the detected radio line emission, they found that the circumstellar SiO fractional abundance (relative to H$_2$) for irregular and semi-regular variables in their sample had a median value of 5$\times$10$^{-6}$,
a factor of about ten lower than expected from theory \citep[see, e.g., review by][]{Millar03}.
 For the high-mass-loss-rate Miras in their sample Gonzalez Delgado et al.\  found a distinct division
into a low-abundance group (on average 4$\times$10$^{-7}$) and a high-abundance group (on average 7$\times$10$^{-6}$). They  concluded that this division is not an
effect of the modelling, but they found no obvious astrophysical 
explanation. 

\begin{table*}
\caption{Integrated ($I_{\mathrm{obs}} = \int T_{\mathrm{mb}}\,dv$) line intensities in K\,km\,s$^{-1}$ for the new observations of SiO ($v$\,$=$\,0, $J$\,$\rightarrow$\,$J-1$) line emission.}
\label{intensities}
$$
\begin{array}{p{0.15\linewidth}cccccccccccc}
\hline
\noalign{\smallskip}
\multicolumn{1}{c}{{\mathrm{Source}}^a} &
\multicolumn{1}{c}{{\mathrm{OSO}}} & &
\multicolumn{2}{c}{{\mathrm{NRAO}}} & &
\multicolumn{3}{c}{{\mathrm{SEST}}}  &&
\multicolumn{1}{c}{{\mathrm{JCMT}}}  &&
\multicolumn{1}{c}{{\mathrm{APEX}}}   \\ 
\cline{2-2}
\cline{4-5}
\cline{7-9}
\cline{11-11}
\cline{13-13}
&
\multicolumn{1}{c}{2\rightarrow1} & &
\multicolumn{1}{c}{3\rightarrow2} &
\multicolumn{1}{c}{5\rightarrow4}& &
\multicolumn{1}{c}{2\rightarrow1} &
\multicolumn{1}{c}{3\rightarrow2} &
\multicolumn{1}{c}{5\rightarrow4} &&
\multicolumn{1}{c}{8\rightarrow7} &&
\multicolumn{1}{c}{8\rightarrow7}\\
\noalign{\smallskip}
\hline
\noalign{\smallskip}
\object{LP And}       & \phantom{0}0.94  && \phantom{0}0.62 & 2.76 && \cdots & \cdots & \cdots && \cdots && \cdots\\
\object{RV Aqr}       & \cdots  && \cdots & \cdots && \cdots & \cdots & \cdots && \cdots && \phantom{0}3.87\phantom{:}\\
\object{U Cam}       & \cdots  && \cdots & \cdots && \cdots & \cdots & \cdots  && 1.20 &&  \cdots\\
\object{S Cep}        & \cdots  && \phantom{0}1.26 & \cdots && \cdots & \cdots & \cdots && \cdots && \cdots \\
\object{V Cyg}        & \cdots &&\phantom{0}1.21 & 4.26 && \cdots & \cdots & \cdots  && \cdots && \cdots\\
\object{R For}       & \cdots  && \cdots & \cdots && \phantom{0}0.32 ^a & \phantom{0}1.00 & \phantom{0}2.72 && \cdots &&  \phantom{0}2.62\phantom{:} \\
\object{V821 Her}   & \cdots  && \cdots & \cdots && \cdots & \cdots & \cdots && \cdots &&  \phantom{0}4.97\phantom{:}\\
\object{CW Leo}    & \cdots &&27.9\phantom{0} & \cdots && 21.7\phantom{0}\phantom{^a}  & 48.4\phantom{0} & 84.4\phantom{0} && \cdots && \cdots \\
\object{R Lep}       & \cdots  && \cdots & \cdots && \phantom{0}0.39 ^a & \phantom{0}0.99 & \phantom{0}3.26  && \cdots &&  \phantom{0}2.73\phantom{:}\\
\object{RW LMi}    & \phantom{0}3.45 && \cdots & \cdots && \phantom{0}1.82\phantom{^a}  &  \phantom{0}4.49 &  \cdots && \cdots && \cdots\\
\object{V384 Per}  & \cdots && \phantom{0}0.75 & \cdots && \cdots &  \cdots &  \cdots  && \cdots \\
\object{W Pic}   & \cdots  && \cdots & \cdots && \cdots & \cdots & \cdots && \cdots &&  \phantom{0}0.94$:$\\
\object{R Vol}       & \cdots  && \cdots & \cdots && \phantom{0}0.67 ^a & \phantom{0}1.04& \cdots  && \cdots && \cdots\\
\object{IRAS 07454--7112}       & \cdots  && \cdots & \cdots && \cdots & \cdots & \cdots && \cdots && \phantom{0}5.12\phantom{:}\\
\object{IRAS 15082--4808}       & \cdots  && \cdots & \cdots && \cdots & \cdots & \cdots  && \cdots && \phantom{0}8.51\phantom{:}\\
\object{IRAS 15194--5115}       & \cdots  && \cdots & \cdots && \cdots & \cdots & \cdots && \cdots &&  20.4\phantom{0}\phantom{:}\\

\noalign{\smallskip}
\hline
\end{array}
$$
$^a$ A colon ($:$) marks a low S/N detection.\\
$^b$ Improved S/N-ratio compared to previous detection by \citet{Olofsson98b}.

\end{table*}

Interestingly, \citet{Delgado03b} found a trend of decreasing circumstellar SiO fractional abundance with
increasing mass-loss rate of the star, interpreted as an 
effect of increased adsorption of SiO onto dust grains with increasing mass-loss
rate, i.e., the density of the envelope. This claim has been further corroborated by the recent
 interferometric SiO observations of the two low-mass-loss-rate M-type AGB stars \object{R~Dor} and \object{L$^2$~Pup}
 performed  by \citet{Schoeier04b}. The excitation analysis of these SiO ($v$=0, $J$=2$\rightarrow$1) 
 interferometric observations 
 requires the fractional abundance of SiO to be high, $\approx$\,4\,$\times$\,10$^{-5}$, within $\approx$\,1\,$\times$\,10$^{15}$\,cm as 
suggested by stellar atmosphere models. Beyond this radius the SiO fractional abundance 
drops by about an order of magnitude and at $\approx 3\times 10^{15}$ cm 
photodissociation destroys the SiO molecules. 
Thus, there exists strong indications that the circumstellar SiO line
emission carries information on the region where the
mass loss is initiated, and where dust formation takes
place.

In this paper we present new (sub-)millimetre line observations of SiO for a sample of carbon stars. The observations are supplemented by a detailed non-LTE radiative transfer analysis in order to obtain reliable circumstellar molecular abundances. The results are then compared to predictions from available chemical models and to the abundance estimates of SiO in M-type (oxygen-rich) AGB stars.

\section{Observations}
\label{sect_obs}
\begin{figure*}
   \centering{   
   \includegraphics[width=16cm]{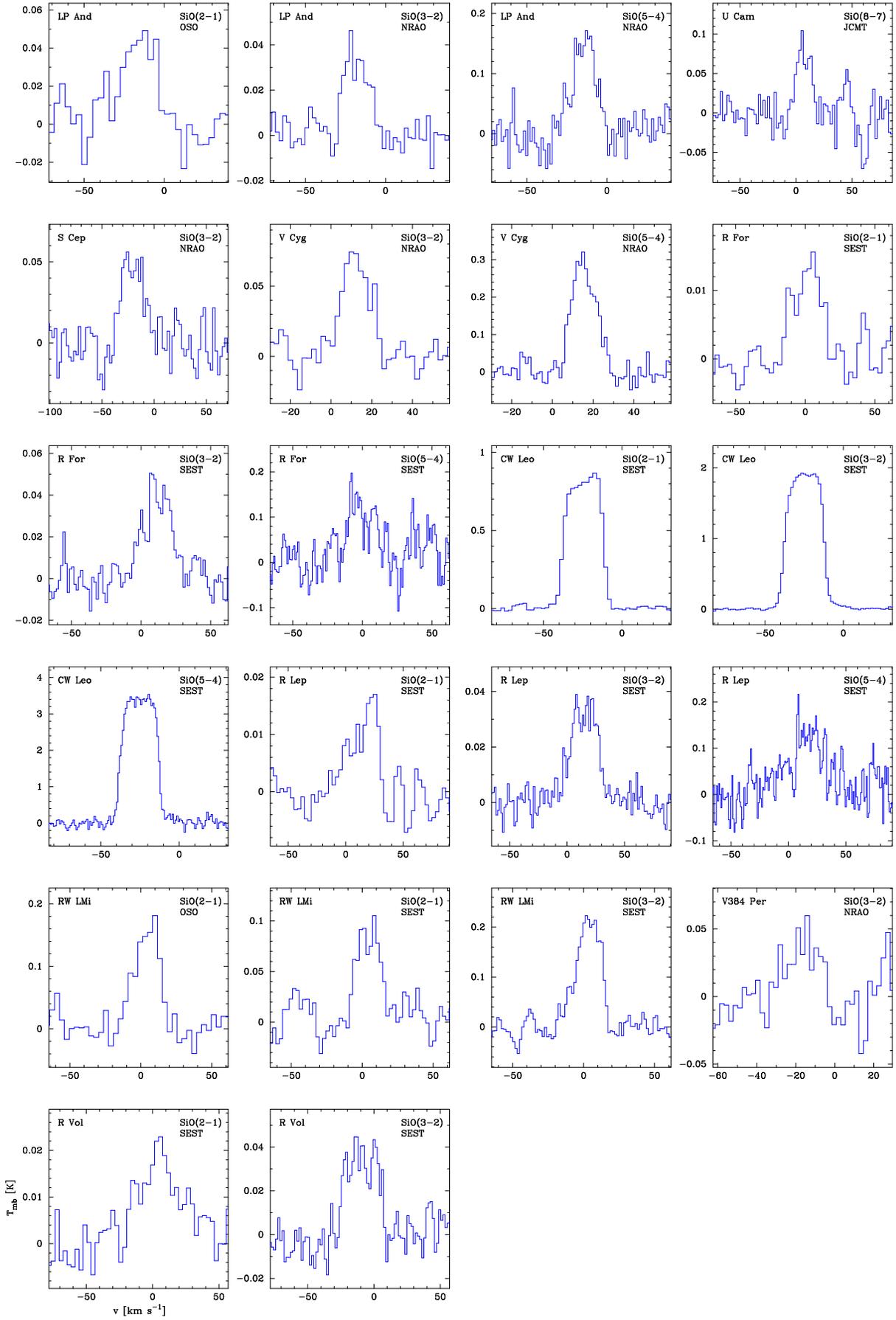}
   \caption{New observations of SiO line emission.}
   \label{obs}}
   \end{figure*}

\begin{figure*}
   \centering{   
   \includegraphics[width=16cm]{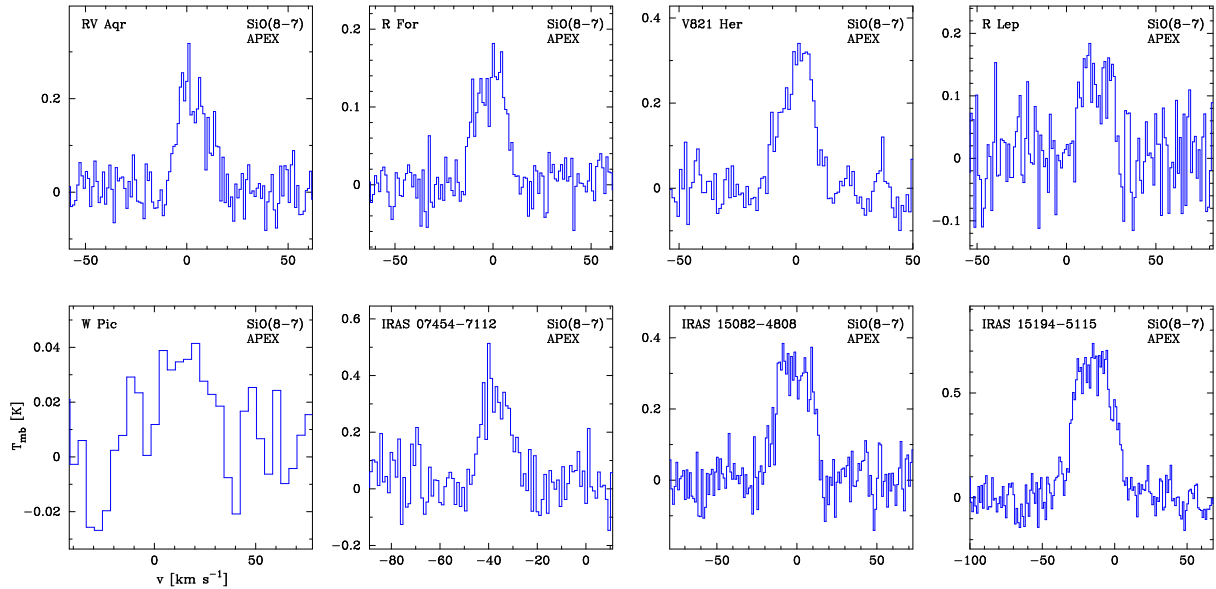}
   \caption{New observations of SiO $J$\,$=$\,8$\rightarrow$\,7 line emission using the APEX telescope. The velocity resolution is 1.0\,km\,s$^{-1}$ in all the spectra, except for \object{W Pic} which has been reduced to a resolution of 4.0\,km\,s$^{-1}$.}
   \label{obs_apex}}
   \end{figure*}

The observations were performed during 1994--1998 using the Onsala 20\,m telescope\footnote{The Onsala 20\,m telescope is operated by the  Swedish National Facility for Radio Astronomy, Onsala Space observatory at Chalmers University of technology} (OSO), the Swedish-ESO submillimetre telescope\footnote{The SEST was located on La Silla, Chile and operated jointly by the Swedish National Facility for Radio Astronomy and the European Southern Observatory (ESO).} (SEST), the NRAO 12\,m telescope at Kitt Peak, in October 2003 using the JCMT telescope\footnote{Based on observations obtained with the James Clerk Maxwell Telescope, which is operated by the Joint Astronomy Centre in Hilo, Hawaii on behalf of the parent organizations PPARC in the United Kingdom, the National Research Council of Canada and The Netherlands Organization for Scientific Research}, and in September 2005 using the APEX telescope\footnote{This publication is based on data acquired with the Atacama Pathfinder Experiment (APEX). APEX is a collaboration between the Max-Planck-Institut fŸr Radioastronomie, the European Southern Observatory, and the Onsala Space Observatory}.
The observed spectra are presented in Figs~\ref{obs} \& \ref{obs_apex} and velocity-integrated intensities are reported in Table~\ref{intensities}. The intensity
scales are given in main-beam brightness temperature scale ($T_{\mathrm{mb}}$).

\begin{table}
\caption{Telescope data.}
\label{efficiencies}
$$
\begin{array}{cccccccccc}
\hline
\noalign{\smallskip}
\multicolumn{1}{c}{{\mathrm{Transition}}} & &
\multicolumn{1}{c}{{\mathrm{Frequency}}} &  &
\multicolumn{1}{c}{{\mathrm{Telescope}}}  &&
\multicolumn{1}{c}{\eta_{\mathrm{mb}}}  && 
\multicolumn{1}{c}{\theta_{\mathrm{mb}}} \\ 
& & 
\multicolumn{1}{c}{{\mathrm{[GHz]}}} && &&  &&
\multicolumn{1}{c}{[\arcsec]} \\
\noalign{\smallskip}
\hline
\noalign{\smallskip}
J=2\rightarrow1 &&  \phantom{0}86.847  &&  \mathrm{SEST}    && 0.75 && 57 \\
                                                      &&                                        &&  \mathrm{OSO}    && 0.65 && 44 \\
                        J=3\rightarrow2 &&                      130.269  &&  \mathrm{NRAO}  && 0.80 && 49 \\
                                                      &&                                       &&  \mathrm{SEST}   && 0.68 && 38 \\
                        J=5\rightarrow4 &&                      217.105  &&  \mathrm{NRAO}  && 0.55 && 29 \\
                                                     &&                                       &&  \mathrm{SEST}    && 0.50 && 24 \\
                                                                                                          &&                                       &&  \mathrm{JCMT}   && 0.70 && 23 \\
                        J=8\rightarrow7 &&                     347.331   &&  \mathrm{APEX}  && 0.70 && 18 \\
                                                     &&                                        && \mathrm{JCMT}   && 0.62 && 14\\

\noalign{\smallskip}
\hline
\end{array}
$$
\end{table}

The SEST, OSO and JCMT observations were made in a dual beamswitch mode, 
where the source is alternately placed in the signal and the reference
beam, using a beam throw of about $11\arcmin$ (SEST and OSO) or $2\arcmin$ (JCMT). 
This method produces very flat baselines. 
At the APEX 12\,m telescope the observations were carried out using a position-switching mode, with the reference position located +2$\arcmin$ in azimuth. 
The raw spectra are stored in the $T_{\mathrm A}^{\star}$  scale and converted to main-beam brightness temperature using $T_{\mathrm{mb}}$\,=\,$T_{\mathrm
A}^{*}/\eta_{\mathrm{mb}}$. $T_{\mathrm A}^{\star}$ is the
antenna temperature corrected for atmospheric attenuation using the
chopper-wheel method, and $\eta_{\mathrm{mb}}$ is the main-beam
efficiency. Regular pointing checks were made on SiO masers (SEST and OSO) and strong CO sources (APEX) and typically found to be consistent within $\approx$\,3$\arcsec$.


At the NRAO 12\,m telescope the observations were carried out using a position-switching mode, with the reference position located +10$\arcmin$ in azimuth. This is the preferred observation mode for spectral line observations at the 12 m telescope. The raw spectra, which are stored in the $T_{\mathrm{R}}^*$ scale, were converted using $T_{\mathrm{mb}}$=$T^*_{\mathrm R}$/$\eta^*_{\mathrm m}$, where $\eta^*_{\mathrm m}$,
is the corrected main-beam efficiency. The  $T_{\mathrm{R}}^*$ scale is related to  $T_{\mathrm{A}}^*$ through  $T_{\mathrm{R}}^*$=$T_{\mathrm{R}}^*$/$\eta_{\mathrm{fss}}$, where $\eta_{\mathrm{fss}}$ is the forward-scattering and spillover efficiency. Regular pointing checks were made on  strong continuum sources  and typically found to be consistent within $\approx$\,5$\arcsec$.

 The adopted beam efficiencies, together with the FWHM of the main beam ($\theta_{\mathrm{mb}}$), for all telescopes and frequencies are given in Table~\ref{efficiencies}.
The uncertainty in the absolute intensity scale is estimated to be about $\pm 20$\%. 

The data was reduced in a standard way, by removing a low order baseline and then binned in order to improve the 
signal-to-noise ratio, using XS\footnote{XS is a package developed by P. Bergman to reduce and analyse a large number of single-dish spectra. It is publically available from {\tt ftp://yggdrasil.oso.chalmers.se}}.

In addition to the new data presented here we have also used SiO line intensities reported by Olofsson et al.\ (1982, 1998)\nocite{Olofsson82} \nocite{Olofsson98b} \& \citet{Woods03} ($J$=2$\rightarrow$1), \citet{Bujarrabal94}  ($J$=2$\rightarrow$1 and 3$\rightarrow$2), and \citet{Bieging00} ($J$=5$\rightarrow$4 and 8$\rightarrow$7). In total, 19 carbon stars have been detected in SiO line emission. 

These stars detected in SiO millimetre line emission provide our sample and are listed in Table~\ref{sample}.

\begin{table*}
\caption{Model results.}
\label{sample}
$$
\begin{array}{p{0.14\linewidth}cccccccccccccccccccccc}
\hline
\noalign{\smallskip}
& 
\multicolumn{8}{c}{\mathrm{SED\ modelling}} & &
\multicolumn{4}{c}{\mathrm{CO\ modelling}} & &
\multicolumn{4}{c}{\mathrm{SiO\ modelling}} \\
\noalign{\smallskip}
\cline{2-9}\cline{11-14}\cline{16-19}
\noalign{\smallskip}
\multicolumn{1}{c}{{\mathrm{Source}}} &
\multicolumn{1}{c}{D}& 
\multicolumn{1}{c}{L_{\star}}&
\multicolumn{1}{c}{T_{\star}}&
\multicolumn{1}{c}{\tau_{10}}&
\multicolumn{1}{c}{T_{\mathrm{c}}}&
\multicolumn{1}{c}{r_{\mathrm{i}}}&
\multicolumn{1}{c}{\chi^2_{\mathrm{red}}} &
\multicolumn{1}{c}{N} & &
\multicolumn{1}{c}{\dot{M}}&
\multicolumn{1}{c}{v_{\mathrm{e}}} &
\multicolumn{1}{c}{\chi^2_{\mathrm{red}}} &
\multicolumn{1}{c}{N} & &
\multicolumn{1}{c}{f_0} & 
\multicolumn{1}{c}{r_{\mathrm{e}}} &
\multicolumn{1}{c}{\chi^2_{\mathrm{red}}} &
\multicolumn{1}{c}{N} \\
&
\multicolumn{1}{c}{[\mathrm{pc}]}  &
\multicolumn{1}{c}{[\mathrm{L_{\odot}}]} & 
\multicolumn{1}{c}{[\mathrm{K}]} & &
\multicolumn{1}{c}{[\mathrm{K}]} &
\multicolumn{1}{c}{[\mathrm{cm}]} & & & &
\multicolumn{1}{c}{[\mathrm{M_{\odot}\,yr^{-1}}]} &
\multicolumn{1}{c}{[\mathrm{km\,s^{-1}}]} & & 
&&&
\multicolumn{1}{c}{[\mathrm{cm}]} & &  \\
\noalign{\smallskip}
\hline
\noalign{\smallskip}
\object{LP And}                       &\phantom{0}630 & 9400 & 2000 & 0.60\phantom{0} & 1100 & 1.8\times10^{14}& 0.8 & 11 && 1.5\times10^{-5} & 13.5 & 0.7 & 7  && 1.2\times10^{-7}\phantom{:} & 2.2\times10^{16} & 2.9 & 3  \\
\object{RV Aqr}                      & \phantom{0}670 & 6800 & 2200 & 0.27\phantom{0} & 1300 & 7.6\times10^{13} & 0.8 & 9  & & 2.8\times10^{-6} & 15.0 & 0.3 & 3 && 4.0\times10^{-6}\phantom{:} & 9.3\times10^{15} & 0.1 & 2\\
\object{UU Aur}                       & \phantom{0}260 & 6900 & 2800 & 0.017 & 1500  &  6.3\times10^{13}& 1.3 & 9 && 2.4\times10^{-7} &10.5& 1.0 & 5  && 1.0\times10^{-6}$:$ & 3.4\times10^{15} & \cdots & 1 \\
\object{U Cam}$^{\mathrm{a}}$ & \phantom{0}340 & 7000  & 2700 & 0.01\phantom{0} &  1500 & 4.4\times10^{13}& 1.5 & 9 && 2.0\times10^{-7} & 11.5 & \cdots & 4 && 5.0\times10^{-5}\phantom{:} & 3.0\times10^{15} & \cdots & 1  \\
\object{S Cep}                        & \phantom{0}380 & 7300 & 2200 & 0.12\phantom{0} & 1400 & 5.8\times10^{13} & 1.5 & 9 && 1.2\times10^{-6} & 21.5 & 0.9 & 5 && 9.0\times10^{-6}\phantom{:} & 4.8\times10^{15} & 0.4 & 4 \\
\object{V Cyg}                         & \phantom{0}310 & 6300 & 1900 & 0.08\phantom{0} & 1200 & 8.7\times10^{13}& 0.6 & 8 && 9.0\times10^{-7} & 10.5 & 0.5 & 5 && 3.5\times10^{-6}\phantom{:} & 6.4\times10^{15} & 3.0 & 6 \\
\object{V1965 Cyg}                &                    1200 & 9800 & 2000 & 0.55\phantom{0} & 1000 & 2.2\times10^{14}& 1.1 & 10 && 1.0\times10^{-5} & 27.0& 2.0 & 3  && 7.0\times10^{-6}\phantom{:} & 1.4\times10^{16} &  \cdots & 1 \\
\object{R For}                          &\phantom{0}610 & 5800 & 2000 & 0.25\phantom{0} & 1400 &  5.6\times10^{13}& 2.8 & 9 && 1.1\times10^{-6} & 16.0 & 1.5 & 7 && 8.0\times10^{-6}\phantom{:} & 5.8\times10^{15} & 0.3 & 4  \\
\object{V821 Her}                   & \phantom{0}600 & 7900 & 2200 & 0.45\phantom{0} & 1500 & 8.1\times10^{13}& 2.4 & 10  && 1.8\times10^{-6} & 13.0 & 3.9 & 4  && 6.0\times10^{-6}\phantom{:} & 8.1\times10^{15} & 0.1 & 2 \\
\object{CW Leo}                     & \phantom{0}120 & 9600 & 2000 & 0.90\phantom{0} & 1200 & 1.7\times10^{14}& 2.1 & 9 && 1.5\times10^{-5} & 14.0 & 0.5 & 8 && 2.0\times10^{-7}\phantom{:} & 2.2\times10^{16} & 1.0 & 9 \\
\object{R Lep}                         & \phantom{0}250 & 4000 & 2200 & 0.06\phantom{0} &  1500 & 4.3\times10^{13}& 0.3 & 9 && 5.0\times10^{-7} & 16.5 & 0.7 & 6 && 2.8\times10^{-6}\phantom{:} & 4.5\times10^{15} & 0.9 & 4  \\
\object{RW LMi}                      & \phantom{0}440 & 9700 & 2000 & 0.50\phantom{0} & 1000 & 2.1\times10^{14}& 1.4 & 11 && 6.0\times10^{-6} & 16.5 & 1.1 & 7 && 2.0\times10^{-6}\phantom{:} & 1.3\times10^{16} & 1.5 & 5  \\
\object{V384 Per}                   & \phantom{0}560 & 8100 & 2000 & 0.25\phantom{0} & 1300 & 1.0\times10^{14}& 1.5 & 11  && 3.5\times10^{-6} & 14.5 & 0.7 & 6 && 1.5\times10^{-6}\phantom{:} & 1.1\times10^{16} & 3.2 & 3 \\
\object{W Pic}                          &   \phantom{0}490  & 4000 & 2500 & 0.015  & 1500  & 4.5\times10^{13} &  3.2  & 8  && 2.3\times10^{-7} & 15.0& 3.4 & 3  && 1.9\times10^{-5}$:$ & 2.8\times10^{15} & \cdots & 1 \\
\object{R Vol}                          &\phantom{0}730 & 6800 & 2000 & 0.30\phantom{0} & 1500 & 6.6\times10^{13}& 1.1 &  9 && 1.7\times10^{-6} & 16.5 & 0.6 & 3 && 9.0\times10^{-6}\phantom{:} & 7.0\times10^{15} & 1.5 & 2  \\
\object{AFGL 3068}                & \phantom{0}980 & 7800 & 2000 & 2.70\phantom{0} & 1100 & 2.5\times10^{14}& 1.8 & 8  && 1.0\times10^{-5} & 13.5 & 0.6 & 4  && 2.0\times10^{-7}\phantom{:} & 1.8\times10^{16} & \cdots & 1 \\
\object{IRAS\,07454--7112}  &\phantom{0}710 & 9000 & 2100 & 0.45\phantom{0} & 1200 & 1.4\times10^{14}&  1.0 & 9 && 5.0\times10^{-6} & 12.5 & 0.1 & 2 && 1.3\times10^{-6}\phantom{:} & 1.3\times10^{16} & 0.3  & 2  \\
\object{IRAS\,15082--4808}  &\phantom{0}640 & 9000 & 2200 & 0.80\phantom{0} & 1100 & 1.9\times10^{14}& 7.0 & 9 && 1.0\times10^{-5} &19.0 & 0.2 & 2  && 1.8\times10^{-6}\phantom{:} & 1.5\times10^{16} &  2.3 & 2\\
\object{IRAS\,15194--5115}  & \phantom{0}500 & 8800 & 2400 & 0.55\phantom{0} & 1200 & 1.5\times10^{14}& 0.4 &  9 && 9.0\times10^{-6} & 21.0 & 0.9 & 4   && 5.0\times10^{-6}\phantom{:} & 1.4\times10^{16} & 1.9 & 2 \\

\noalign{\smallskip}
\hline
\end{array}
$$
\noindent
$^{\mathrm{a}}$  U Cam has a detached shell that complicates the analysis of the present-day mass-loss characteristics \citep[for details see][]{Schoeier05b}.\\
\noindent
A colon (:) marks an uncertain abundance estimate.
\end{table*}

\section{Molecular excitation analysis}
\label{sect_model}
\subsection{Radiative transfer model}
The CSEs are assumed to be spherically symmetric, produced by a constant mass loss rate ($\dot{M}$), and to expand at a constant velocity ($v_{\mathrm e}$). 
%
%
In order to determine the molecular excitation in the CSEs a detailed non-LTE radiative transfer code, based on the Monte Carlo method,
was used.  The code is described in detail in \citet{Schoeier01} and has been
benchmarked, to high accuracy, against a wide variety of molecular-line radiative
transfer codes in \citet{Zadelhoff02} and van~der~Tak et al.\ (in prep.).  

\citet{Schoeier01} set out to determine the circumstellar physical properties, such as the density, temperature, and kinematic structures, of a large sample of carbon stars based on radiative transfer modelling of millimetre CO line observations. These models, and subsequent models presented in \citet{Schoeier02b} and \citet{Woods03}, form the basis of the underlying physical structure and are used as input to the SiO excitation analysis. In a similar way new circumstellar models where obtained for \object{V821 Her} (\object{IRAS 18397+1738}) and \object{V1965 Cyg} (\object{IRAS 19321+2757}) using CO millimetre line observations published in \citet{Margulis90}, \citet{Nyman92}, and from the JCMT data archive\footnote{{\tt
http://www.jach.hawaii.edu/JCMT/archive/}.}.

The local line width is assumed to be described by a Gaussian and is made up of a micro-turbulent component and a thermal component.
The micro-turbulence is assumed to have a width of 1.0\,km\,s$^{-1}$ \citep{Schoeier04b}.
This is significantly higher than the value of 0.5\,km\,s$^{-1}$ adopted in the CO modelling performed by  \citet{Schoeier01}. However, as pointed out by Sch\"oier \& Olofsson the exact value adopted for the micro-turbulence only has a minor effect on the derived mass-loss rates, as long as it is much lower than the expansion velocity of the wind. The (sub)millimetre SiO observations typically probe regions of gas closer to the star than does the CO sub(millimetre) observations. Observations of IR-absorption lines \citep{Monnier00} indicate that most likely there is a gradient in turbulent velocity, with regions closer to the photosphere being more turbulent. Thus, it is not unreasonable to adopt a higher turbulent velocity when modelling the SiO line emission. The thermal line broadening is calculated from the derived kinetic temperature structure.

Since the publication of \citet{Schoeier01} new collisional rate coefficients of CO have been calculated by \citet{Flower01a}. These new rate coefficients were adopted here and extended to include more energy levels as well as extrapolated in temperature, as described in \citet{Schoeier05a}. An ortho-to-para ratio of 3.0 was adopted when weighting together collisional rate coefficients for CO in collisions with ortho-H$_2$ and para-H$_2$. In addition, dust emission has been included in the excitation analysis for CO in the same fashion as for SiO described below. Moreover, new CO observations of $J$\,$=$\,3\,$\rightarrow$\,2 line emission using the JCMT and APEX telescopes (Sch\"oier \& Olofsson, in prep.) has been included in the modelling. In all, these new modifications to the excitation analysis of CO millimetre line emission have not altered the mass-loss rates derived by \citet{Schoeier01} by more than 20\%, i.e., within the calibration uncertainty of the observations. The new mass-loss rates are reported in Table~\ref{sample}.

The excitation analysis includes radiative excitation through the first vibrationally excited ($v$\,=\,1) state for both CO at 4.6\,$\mu$m and SiO at 8\,$\mu$m. Relevant molecular data are summarized in \citet{Schoeier05a} and are made publicly available through the {\em Leiden Atomic and Molecular Database} (LAMDA){\footnote{\tt http://www.strw.leidenuniv.nl/$\sim$moldata}}. The majority of sources in this study are intermediate- to high-mass-loss-rate objects where thermal dust emission provides the main source of infrared photons which excite the $v$\,=\,1 state.

The addition of a dust component in the Monte Carlo scheme is straightforward as described in \citet{Schoeier02b}. At the wavelengths of importance here ($\lambda$\,$\geq$\,4.8\,$\mu$m)
scattering can be neglected and only emission and absorption by the dust particles need to be considered. 
The number of model photons emitted per second by the dust within a small volume of the envelope, $\Delta V$ , and over a small frequency interval, 
$\Delta\nu$, can then be written as 
\begin{equation}
N_{\mathrm{d}} (r) = \frac{4\pi}{h\nu}\kappa_{\mathrm{\nu}}\rho_{\mathrm{d}}(r)B_{\nu}[T_{\mathrm{d}}(r)]\Delta V(r)\Delta\nu,
\end{equation}
where $\kappa_\nu$ is the dust opacity per unit mass, $\rho_{\mathrm{d}}$ the dust mass density, 
and $B_\nu$ the Planck function. $T_{\mathrm{d}}(r)$ is the temperature of the thermally emitting dust grains.
Typically, the frequency passband corresponds to 3$\times$$v_{\mathrm{e}}$ and it is centered on the line rest frequency $\nu_0$. 

The model photons emitted by the dust are released together with the other 
model photons. The additional opacity provided by the dust, 
\begin{equation}
\tau_{\nu} = \kappa_{\mathrm{\nu}}\int \rho_{\mathrm{d}}(r)\,dr,
\end{equation}
is added to the line optical depth. 

\subsection{SED modelling}
The dust-temperature structure and dust-density profile are obtained from detailed 
radiative transfer modelling using {\em Dusty} \citep{Ivezic97}. 
Here we are only interested in the potential of dust grains to affect the excitation of molecules. 
A full treatment of dust radiative transfer coupled with a dynamical model for the wind, in order to independently measure physical 
properties of the CSE such as, e.g., the mass-loss rate, is beyond the scope of this paper and left to a future publication (Sch\"oier \& Olofsson, in prep.).
In the dust modelling the same basic assumptions are made 
as for the gas modelling, i.e., a spherically symmetric envelope 
expanding at a constant velocity. 
Prompt dust formation is assumed at the inner  radius, $r_{\mathrm{i}}$, of the CSE.
Amorphous carbon dust grains 
with the optical constants given in \citet{Suh00} are adopted. For simplicity, the 
dust grains are assumed to be of the same size (a radius, $a_{\mathrm{d}}$, of 
0.1\,$\mu$m), and the same mass density ($\rho_{\mathrm{s}}$\,=\,2.0\,g\,cm$^{-3}$). 
The 
corresponding dust opacities, $\kappa_{\nu}$, were then calculated from the optical constants 
and the individual grain properties using standard Mie theory 
\citep{Bohren83}. 

\begin{figure}
   \centering{   
   \includegraphics[width=7cm]{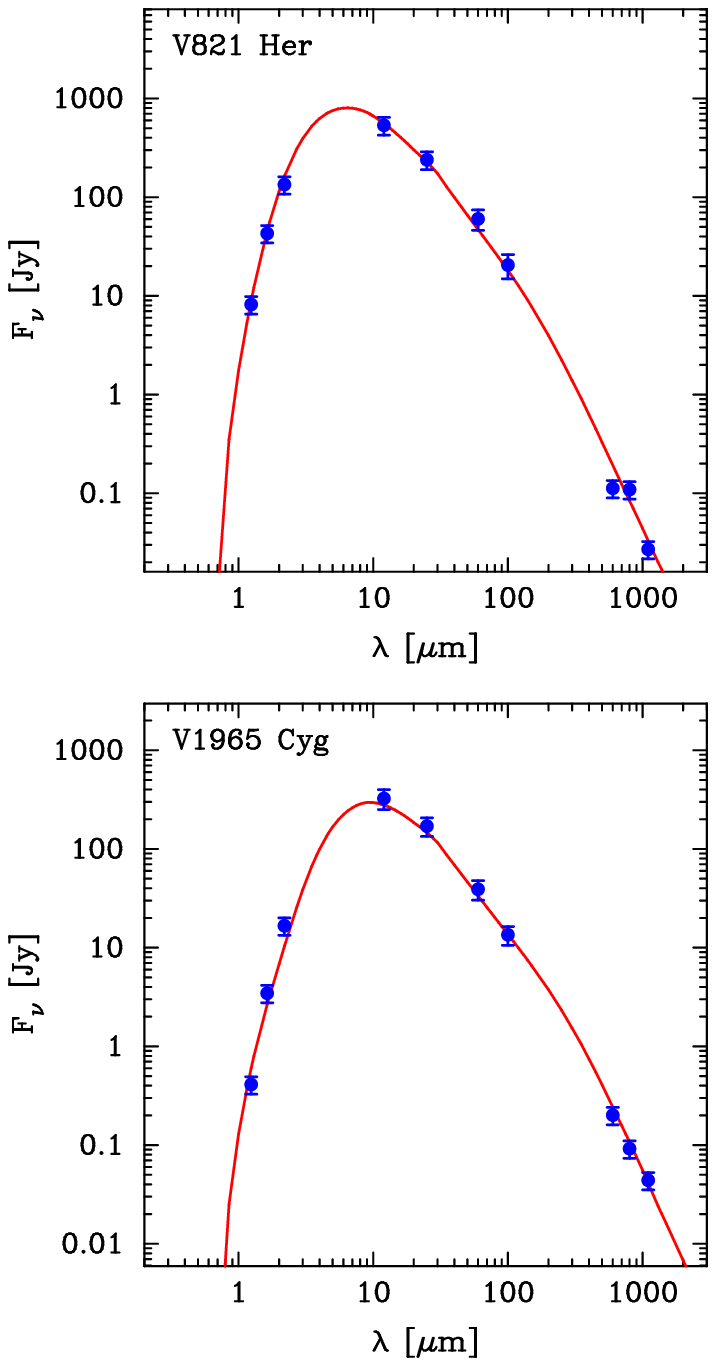}
   \caption{Best-fit models,  derived from the dust radiative transfer modelling (parameters given in Table~\ref{sample}), of the observed SEDs for \object{V821 Her} and \object{V1965 Cyg} .}
   \label{sed}}
   \end{figure}

In the modelling, where the SED provides the observational 
constraint, the dust optical depth specified at 10\,$\mu$m, $\tau_{10}$, and 
the dust condensation temperature, $T_{\mathrm{d}} (r_{\mathrm{i}})$, 
are the adjustable parameters in the $\chi^2$-analysis. The model SED only weakly depends 
on the other input parameters which are fixed at reasonable values. The SED typically consists of {\em{JHKLM}} photometric data \citep[][ and Kerschbaum priv. com.]{Kerschbaum99b}, IRAS fluxes, and in some cases sub-millimetre data \citep{Groenewegen93}.
The total luminosity of the source is obtained from the period-luminosity relation of \citet{Groenewegen96}. The distance is then obtained from the SED fitting.
The derived parameters are reported in Table~\ref{sample} and examples of best-fit models are shown in Fig.~\ref{sed} for  the new sample sources \object{V821~Her} and \object{V1965~Cyg}. 
A more detailed description of the procedure as well as other examples for the sample stars can be found in \citet{Schoeier02b}.

   \begin{figure*}
   \centering{   
   \includegraphics[width=16cm]{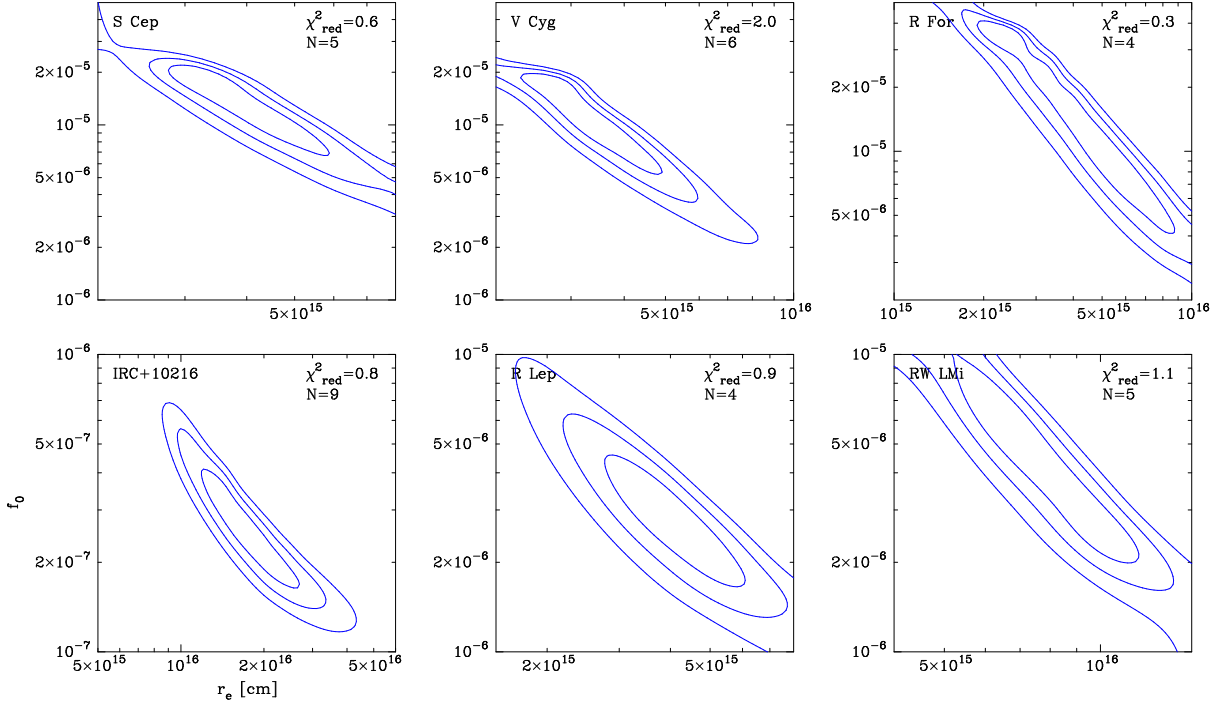}
   \caption{$\chi^2$ maps showing the sensitivity of the excitation analysis to the adopted SiO fractional abundance ($f_0$) and envelope size ($r_{\mathrm e}$). Contours are drawn at the 1, 2, and 3$\sigma$ levels. Also indicated is the lowest reduced $\chi^2$ ($\chi^2_{\mathrm{red}}=\chi^2_{\mathrm{tot}}/(N-2)$) in the map.}
   \label{sio_chi2}}
   \end{figure*}

\subsection{SiO envelope sizes}
In order to determine accurate abundances it is important to know the spatial extent of the molecules. \citet{Delgado03b} showed that it is possible to put constraints also on the size of the SiO line emitting envelope from single-dish observations alone, given that enough transitions are observed (in practice four or more). In the present sample six carbon stars (\object{CW Leo},  \object{R For},  \object{R Lep}, \object{RW LMi}, \object{S Cep}, and \object{V Cyg}) have been observed in four or more transitions. For these sources the size of the SiO emitting region ($r_{\mathrm e}$) and the inner fractional abundance of SiO ($f_0$) are varied simultaneously. The abundance distribution is assumed to be described by a Gaussian
\begin{equation}
\label{eq_distr}
f(r) = f_0\, \exp \left(-\left(\frac{r}{r_{\mathrm e}}\right)^2 \right),
\end{equation}
where $f$\,$=$\,$n\mathrm{(SiO)}/n\mathrm{(H_2)}$, i.e., the ratio of the number density of SiO molecules to that of H$_2$ molecules. In circumstellar enevelopes such as these H is expected to be mainly in molecular form.

The best fit model is found by minimizing the total $\chi^2$ defined as
\begin{equation}
\label{chi2_sum}
\chi^2_{\mathrm{tot}} = \sum^N_{i=1} \left [ \frac{(I_{\mathrm{mod}}-I_{\mathrm{obs}})}{\sigma}\right ]^2, 
\end{equation} 
where $I$ is the integrated line intensity and $\sigma$ the uncertainty in the measured 
value (usually dominated by the calibration uncertainty of $\pm$20\%), and the summation is done over
$N$ independent observations. The sensitivity of the line emission to variations of the two adjustable parameters $r_{\mathrm{e}}$ and $f_0$ is illustrated in Fig.~\ref{sio_chi2}. As can be seen reasonable estimates can be found for all four objects, also the quality of the best-fit models are good, typically $\chi^2_{\mathrm{red}}$\,$=$\,$\chi^2_{\mathrm{tot}}/(N-2)$\,$\sim$\,1. The 1$\sigma$ range of acceptable values for $r_{\mathrm{e}}$ and $f_0$ are reported in Table~\ref{table_chi2}. However, note that the range of acceptable values does not follow a rectangular shape as seen in Fig.~\ref{sio_chi2}. 

\begin{table}
\caption{SiO abundance distribution.}
\label{table_chi2}
$$
\begin{array}{p{0.2\linewidth}cccc}
\hline
\noalign{\smallskip}
&
\multicolumn{1}{c}{{f_0} ^a} & 
\multicolumn{1}{c}{{r_{\mathrm{e}}} ^a} & 
\multicolumn{1}{c}{N}  &
\multicolumn{1}{c}{\chi^2_{\mathrm{red}}}  \\ 

\multicolumn{1}{c}{{\mathrm{Source}}} &
 &
\multicolumn{1}{c}{[{\mathrm{cm}}]} & &\\
\noalign{\smallskip}
\hline
\noalign{\smallskip}
\object{S Cep}        &  1.4\pm0.8\times10^{-5} &  4.4\pm1.6\times10^{15} & 5 & 0.6\\
\object{V Cyg}        &  1.3\pm0.7\times10^{-5} &  3.6\pm1.3\times10^{15} & 6 & 2.0\\
\object{R For}        &  2.2\pm1.8\times10^{-5} &  5.4\pm3.6\times10^{15} & 4 & 0.3\\
\object{CW Leo}    &   2.8\pm1.1\times10^{-7} &  1.9\pm0.7\times10^{16} & 9 & 0.8\\
\object{R Lep}        &  3.1\pm1.4\times10^{-6} &  4.4\pm1.6\times10^{15} & 4 & 0.9\\
\object{RW LMi}    &  6.0\pm4.0\times10^{-6} &  7.7\pm2.5\times10^{15} & 5 & 1.1\\
\noalign{\smallskip}
\hline
\end{array}
$$
$^a$ The abundance distribution is assumed to be Gaussian (see Eq.~\ref{eq_distr})
\end{table}
   \begin{figure}
   \centering{   
   \includegraphics[width=7cm, angle=-90]{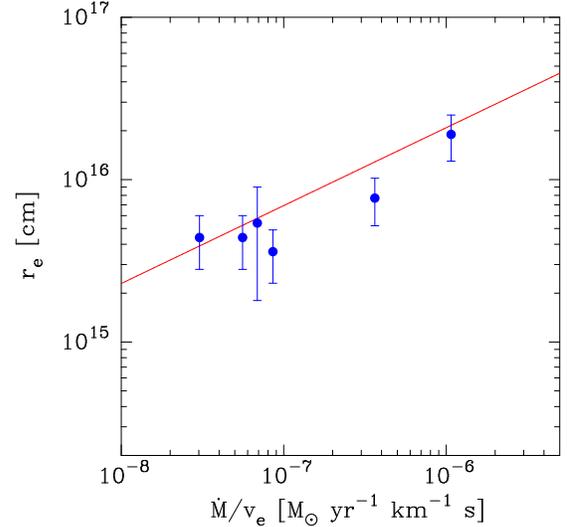}
   \caption{SiO envelope sizes ($r_\mathrm{e}$) estimated from the excitation analysis plotted against a density measure ($\dot{M}$/$v_{\mathrm{e}}$). The solid line is the relation found for a larger sample of M-type AGB stars (Eq.~\ref{eq_size}).}
   \label{sio_size}}
   \end{figure}

Given the relatively low number of carbon stars where $r_{\mathrm e}$ can be determined we have assumed it to scale with density as
\begin{equation}
\label{eq_size}
\log r_{\mathrm{e}} = 19.2 + 0.48 \log \left( \frac{\dot{M}}{v_{\mathrm e}} \right),
\end{equation}
where $\dot{M}$ is the mass-loss rate and $v_{\mathrm e}$ is the expansion velocity of the wind. This scaling law was found by \citet{Delgado03b} for a larger sample of M-type AGB stars. 
As shown in Fig.~\ref{sio_size} this scaling law can, within the uncertainties, account for the estimated envelope sizes derived for the carbon stars in our study.

The radial distribution of SiO molecules in the gas-phase is most probably  
determined by a combination of condensation and photodissociation processes. Thus, one can imagine an initial (pre-condensation) SiO abundance determined by the stellar atmosphere chemistry, possibly modified by non-LTE chemical processes due to shocks in the inner part of the wind out to a few stellar radii \citep[e.g.,][]{Willacy98}. In terms of classical evaporation theory \citep[see][ and references therein]{Delgado03b} the region over which the SiO abundance is expected to change significantly due to condensation onto dust grains is confined within approximately 10$^{15}$\,cm. For intermediate- to high-mass-loss-rate objects this region is significantly smaller than the post-condensation region (see Fig.~\ref{sio_size}), which is limited by photodissociation, and it does not contribute notably to the observed single-dish fluxes. Thus, a single component is adequate for describing the SiO abundance structure in this case. It should be pointed out that this also means that no constraints can be put on the pre-condensation abundance from the present observations. This requires high spatial resolution observations \citep{Schoeier04b}. In the case of low-mass-loss-rate objects the post-condensation region is smaller, but, since the freeze out is less effective, the contrast  between pre- and post-condensation SiO abundances is small. Hence, a single component description is adequate also in this case.

\subsection{SiO abundances}
\begin{figure*}
   \centering{   
   \includegraphics[width=17cm]{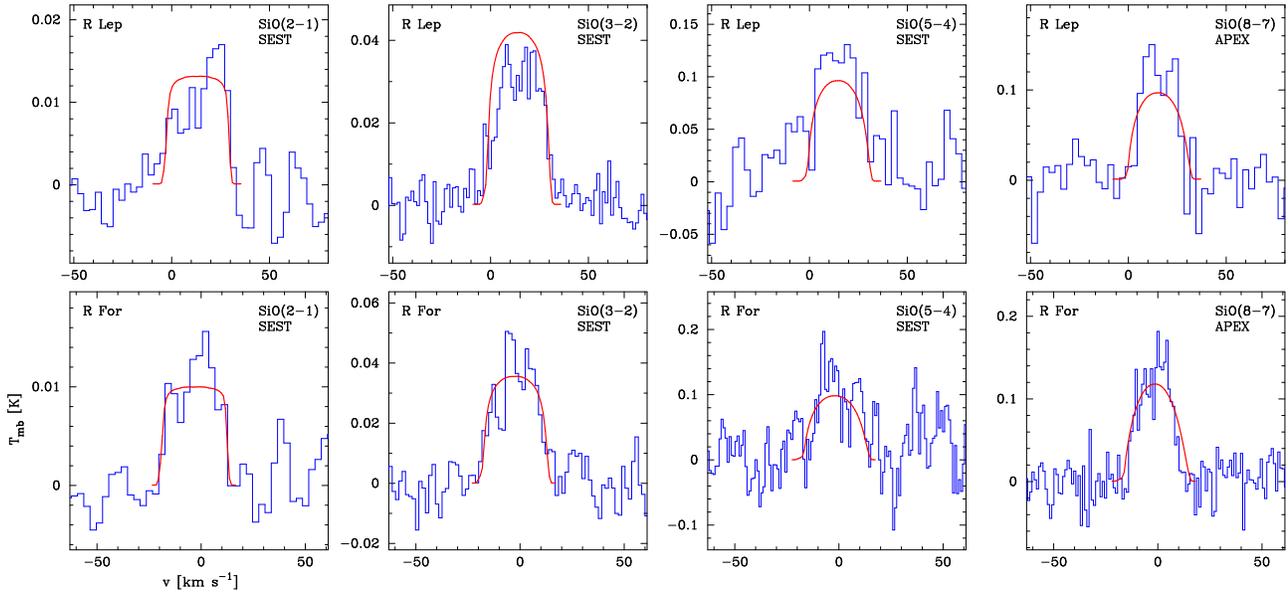}
   \caption{Best-fit models (solid lines; parameters given in Table~\ref{sample}) for \object{R~Lep} and \object{R~For} overlayed on observed spectra (histograms).}
   \label{model}}
   \end{figure*}

Using the scaling relation in Eq.~\ref{eq_size} it is possible to estimate the characteristic size of the envelope ($r_{\mathrm e}$) containing SiO molecules for each of the sample sources.  Once the size is known only one free parameter remains, the fractional abundance of SiO ($f_0$). The derived sizes and abundances are reported in Table~\ref{sample}. The fits are generally excellent with reduced $\chi^2$ values of $\approx$1--2. As examples we show the best-fit
models for \object{R~For} and \object{R~Lep} in Fig.~\ref{model}.

From the modelling it is noted that the SiO $J=3\rightarrow2$ line observed using the IRAM 30\,m telescope by \citet{Bujarrabal94} is consistently lower by about a factor of two compared to model predictions. A possible explanation for this is pointing offsets on the order of $\approx 8\arcsec$.  Due to the larger beam the $J=2\rightarrow1$ observations are less affected, typically on the same level as the absolute calibration uncertainty $\approx 20$\%. However a calibration error can not be ruled out. In the present analysis the SiO $J=3\rightarrow2$  line intensities reported by  \citet{Bujarrabal94} were not included, except for \object{UU~Aur} which is only detected in this line. 

Making an error estimate of the derived abundance is difficult as discussed in \citet{Delgado03b}.  We have also tested the effect of the assumed abundance distribution by replacing the adopted Gaussian distribution (Eq. \ref{eq_distr}) by an exponential decline of the abundance, $f(r)$\,$=$\,$f_0$\,$\exp (-r/r_{\mathrm e})$, as would be the case for a simple photodissociation model without any dust shielding. In the case of a high-mass-loss-rate object such as \object{CW~Leo} the line intensities change by $\lesssim$\,5\% and for a low-mass-loss-rate object such as \object{R~Lep} by $\lesssim$\,20\%. Thus, no significant systematic trends are introduced through the choice of the SiO abundance description.
Typically, we believe the abundances to be accurate to about $\pm$\,50\%, within the adopted circumstellar model.

\subsection{Importance of vibrational excitation}
\label{sect_rovib}
The excitation analysis includes the possibility of populating also the first vibrationally-excited state of SiO through the absorption of 8\,$\mu$m photons, primarily, from dust emission. This can significantly change the derived line intensities of transitions within the ground vibrational state as illustrated in Table~\ref{table_rovib} for \object{CW Leo} (and \object{R~Lep}) for selected transitions. The calculations were performed for a 15\,m telescope. Neglecting to take account of the dust leads to about 50\% lower integrated (over the line) intensities,  which translates into at least a doubled fractional abundance of circumstellar SiO to produce the same flux. Interestingly, in this example approximating the dust emission using a single low-temperature ($T$\,$=$\,510\,K) blackbody containing the total luminosity of the source produces only about 10\% higher intensities compared with the full model taking the dust emission/absorption fully into account. This is not always the case and great caution should be taken when such approximations are made.
It should be noted that \object{CW Leo} is a high-mass-loss-rate object and that the importance of the circumstellar dust radiation in relation to that of the stellar radiation field is less for a lower-mass-loss-rate object such as \object{R~Lep} (Table~\ref{table_rovib}). 

\begin{table}
\caption{Importance of IR pumping in the excitation of SiO.}
\label{table_rovib}
$$
\begin{array}{p{0.25\linewidth}cccc}
\hline
\noalign{\smallskip}
&
\multicolumn{1}{c}{I(2\rightarrow1)} & 
\multicolumn{1}{c}{I(3\rightarrow2)} & 
\multicolumn{1}{c}{I(5\rightarrow4)}  &
\multicolumn{1}{c}{I(8\rightarrow7)}  \\ 

\multicolumn{1}{c}{{\mathrm{Model\ with}}} &
\multicolumn{1}{c}{[{\mathrm{K\,km\,s}}^{-1}]} &
\multicolumn{1}{c}{[{\mathrm{K\,km\,s}}^{-1}]} &
\multicolumn{1}{c}{[{\mathrm{K\,km\,s}}^{-1}]} &
\multicolumn{1}{c}{[{\mathrm{K\,km\,s}}^{-1}]} 
\\
\noalign{\smallskip}
\hline
\noalign{\smallskip}
{\em CW Leo} \\
no (star + dust)       &  11.2 & 22.8 & \phantom{0}41.1 & \phantom{0}74.3 \\
star                           &  13.1 & 28.0 & \phantom{0}46.8 & \phantom{0}81.0\\
star + dust               &  16.5 &  47.3 & \phantom{0}98.9 & 159.2 \\
510\,K BB                &   20.4 &  51.5 & 100.3 & 172.6 \\

\noalign{\smallskip}
{\em R Lep} \\
no (star + dust)       &  0.27 & 0.60 & 1.1 & 2.0 \\
star                           &  0.34 & 0.88   & 1.7 & 2.8 \\
star + dust               &  0.34  & 1.08  & 2.6 & 4.5   \\

\noalign{\smallskip}
\hline
\end{array}
$$
\end{table}

\section{Discussion}
\label{sect_discussion}

\subsection{Comparison with chemical models}
LTE stellar atmosphere models predict  that the SiO fractional abundance in carbon stars is relatively low, typically $\sim$\,5$\times$\,10$^{-8}$ \citep[see reviews by][ and references therein]{Glassgold99, Millar03} about three orders of magnitudes lower than in M-type AGB stars.
The relatively high SiO fractional abundances, derived in Sect.~\ref{sect_model}, of 1$\times10^{-7}$$-$5$\times10^{-5}$ can generally not be explained by LTE chemistry.

Departure from LTE could be caused by the variable nature of AGB stars that induces shocks propagating through the photosphere thereby affecting its chemistry. Models of shocked carbon-rich stellar atmospheres performed by \citet{Willacy98} and \citet{Helling01} indicate that the SiO fractional abundance can be significantly increased by the passage of periodic shocks. There is also a strong dependence on the shock strength and possibly this mechanism can explain the observed fractional abundances, and their large spread, of SiO in carbon stars. 

   \begin{figure}
   \centering{   
   \includegraphics[angle=-90,width=8cm]{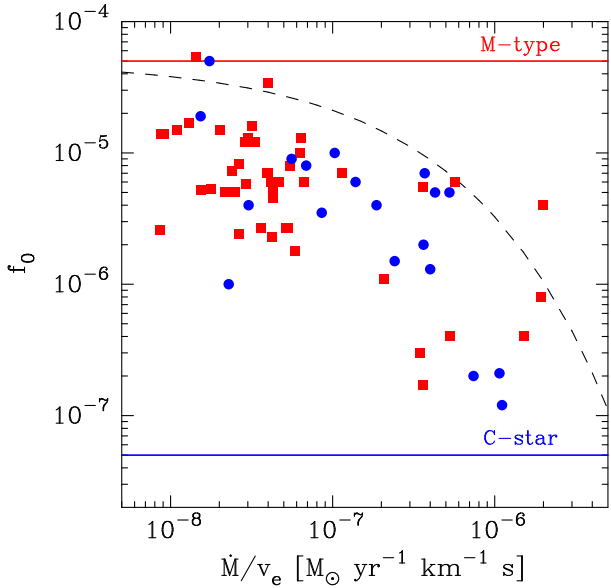}
   \caption{SiO fractional abundance ($f_0$) obtained from an excitation analysis,  as a function of a density measure ($\dot{M}$/$v_{\mathrm{e}}$), for carbon stars (filled circles) and M-type (O-rich) AGB stars (filled squares). The horizontal lines mark the abundances predicted from equilibrium chemistries. The dashed line shows the expected $f(\infty)$ (scaled to 5$\times 10^{-5}$, roughly the expected fractional abundance from stellar equilibrium chemistry when C/O$<$1) for a model including adsorption of SiO onto dust grains \citep[see][ for details]{Delgado03b}.}
   \label{sio_abundance}}
   \end{figure}

From the results reported in Table~\ref{sample} it is clear that the fractional abundance of SiO varies substantially  between the carbon stars in the sample, from as low as $1\times10^{-7}$ up to $5\times10^{-5}$. As illustrated in Fig.~\ref{sio_abundance} (solid circles) there is a clear trend that SiO becomes less abundant as the density, $\dot{M}$/$v_{\mathrm{e}}$, in the wind increases. Even more intriguing, when compared to the distribution of SiO fractional abundances for M-type AGB stars derived by \citet{Delgado03b} (Fig.~\ref{sio_abundance}; filled squares), there appears to be no way of distinguishing a C-rich chemistry from that of an O-rich based on an estimate of the circumstellar SiO abundance alone. Also, the puzzling division of high-mass-loss-rate stars, into distinct low- and high-SiO fractional abundance groups, found by \citet{Delgado03b} for M-type stars is reenforced by the excitation analysis of the carbon stars.

The non-detection of SiO maser emission (e.g., from the $v$\,$=$\,1, $J$\,$=$\,2\,$\rightarrow$\,1 transition) towards carbon stars  \citep[e.g.,][]{Lepine78} is interesting in light of the high circumstellar SiO abundances found. Since SiO maser emission detected from M-type AGB stars emanates very close to the photosphere  and well within the acceleration region \citep[e.g.,][]{Cotton04} this could suggest that in carbon stars the SiO molecules form somewhat further out in the wind where pumping by IR photons, and hence excitation of vibrationally excited states, is less effective.

\subsection{Condensation onto dust grains}
The importance of dust grains in regulating the circumstellar fractional abundance of SiO is supported by the observed trend that the SiO fractional abundance decreases with increasing  density  ($\dot{M}/v_{\mathrm{e}}$) of the wind (Fig.~\ref{sio_abundance}). This trend is expected from a classical treatment of the condensation process, as  described in \citet{Delgado03b}, where the condensation efficiency strongly depends on the dust mass-loss rate.  In Fig.~\ref{sio_abundance} a depletion curve based on the simple model presented in \citet{Delgado03b} is shown. Since the theoretical condensation results are very sensitive to the adopted parameters a large scatter around this curve is natural.

In Fig.~\ref{sio_abundance} the derived abundances are plotted against a density measure where the terminal expansion velocity of the wind has been used. Since much of the condensation may occur in the acceleration region a lower velocity (and hence a higher density) is perhaps more suitable. However, if the characteristic velocity at which dust condensation is close to complete scales with the terminal expansion velocity (independently of the mass-loss rate),  the results would only be shifted along the density axis in Fig.~\ref{sio_abundance}. If, on the other hand, this velocity is independent of the mass-loss rate plotting the abundance against the mass-loss rate is to be preferred. However, this only marginally changes the appearance of  Fig.~\ref{sio_abundance} \citep[see also Fig.~5 in ][]{Delgado03b}.

The inclusion of adsorption of molecules onto dust grains is a shortcoming of current generation chemical models  and will be required in a full model describing the chemistry of AGB stars. In addition, grain surfaces could act as catalysts for chemical reactions. 
Although LTE chemistry can explain the SiO fractional abundances for a few of the carbon stars with the highest mass-loss rates we find it more likely that these abundances are  the result of SiO depletion in the wind and that the photospheric abundance is significantly higher.
High spatial resolution interferometric observations for a sample of sources could help to solidify this claim.

\subsection{The HCN/SiO line intensity ratio}
The HCN/SiO line intensity ratio has been shown to be a useful tool to discriminate between O-rich and C-rich envelopes \citep{Bujarrabal94,Olofsson98b,Bieging00}. The explanation for this has been that in O-rich envelopes the abundance of HCN is very low and that of SiO high, naturally  producing a low HCN/SiO line-intensity ratio. In a C-rich envelope HCN is abundant whereas SiO is not, producing a high HCN/SiO line-intensity ratio. Caution has to be taken here since optical depth effects are present, in particular for high-mass-loss-rate objects.

From the present analysis it is suggested that the SiO fractional abundance in carbon stars closely mimic that in M-type (O-rich) stars. Therefore, we conclude that the observed difference in the HCN/SiO line intensity ratio of about an order of magnitude is mainly due to a significantly lower fractional abundance of HCN in M-type AGB stars compared with the carbon stars. This claim has to be verified by performing a similar excitation analysis, as presented for SiO here and in \citet{Delgado03b},  also for a statistically significant sample of sources detected in HCN line emission. 

\section{Conclusions}
New multi-transition millimetre SiO line observations of a sample of carbon stars are presented. With this addition, a total of 19 carbon stars have been detected in SiO line emission. From a detailed excitation analysis, based on a reliable physical model of the sources, we reach the following conclusions:

\begin{itemize}

\item In order to derive reliable fractional abundances of SiO, excitation through IR ro-vibrational transitions, needs to be taken into account. This will in most cases mean that the effect of dust grains need to be included in the excitation analysis. The present work contains a proper treatment of this.

\item The fractional abundance of SiO in carbon stars  can be several orders of magnitude higher than predicted by thermal equilibrium chemistry. 

\item A possible explanation for the high SiO fractional abundances found is a shock-induced chemistry. However,  the influence of dust grains, both as a source for depletion as well as production of SiO, needs to be further investigated.     

\item As observed for M-type AGB stars, a clear trend that the SiO fractional abundance in carbon stars decreases as the  density ($\dot{M}/v_{\mathrm{e}}$) of the wind increases is found, indicative of adsorption of SiO onto dust grains.



\end{itemize}

 However, these conclusions still
rest on somewhat loose ground, e.g., the constraints on the SiO abundance distribution 
are poor, and the relative importance of freeze-out onto dust grains, photodissociation, and circumstellar chemistry is still uncertain. More observations are required, in particular interferometric observations and high-$J$ single dish observations, to more firmly establish these claims. Also, a larger number of low-mass-loss-rate carbon stars must be observed with high sensitivity. 
 
\begin{acknowledgements}
The authors are grateful to the staff at the APEX telescope.  An anonymous referee is thanked for insightful comments.
FLS and HO acknowledge financial support from the Swedish Research Council.
\end{acknowledgements}

\bibliographystyle{aa}

\end{document}